# A transmission electron microscope study of Néel skyrmion magnetic textures in multilayer thin film systems with large interfacial chiral interaction

S. McVitie[1], S. Hughes[1], K. Fallon[1], S. McFadzean[1], D. McGrouther[1], M. Krajnak[1]*, W. Legrand[2], D. Maccariello[2], S. Collin[2], K. Garcia[2], N. Reyren[2], V. Cros[2], A. Fert[2], K. Zeissler[3], C. H. Marrows[3]

[1]*Scottish Universities Physics Alliance, School of Physics and Astronomy, University of Glasgow, Glasgow G12 8QQ United Kingdom*

[2]*Unité Mixte de Physique, CNRS, Thales, Univ. Paris-Sud, Université Paris-Saclay, 91767 Palaiseau, France*

[3]*School of Physics and Astronomy, University of Leeds, Leeds LS2 9JT, United Kingdom*

*\*Current address: Department of Materials Science and Engineering, Monash University, Clayton, Victoria 3800, Australia.*

Skyrmions in ultrathin ferromagnetic metal (FM)/heavy metal (HM) multilayer systems produced by conventional sputtering methods have recently generated huge interest due to their applications in the field of spintronics. The sandwich structure with two correctly-chosen heavy metal layers provides an additive interfacial exchange interaction which promotes domain wall or skyrmion spin textures that are Néel in character and with a fixed chirality. Lorentz transmission electron microscopy (TEM) is a high resolution method ideally suited to quantitatively image such chiral magnetic configurations. When allied with physical and chemical TEM analysis of both planar and cross-sectional samples, key length scales such as grain size and the chiral variation of the magnetisation variation have been identified and measured. We present data showing the importance of the grain size (mostly < 10nm) measured from direct imaging and its potential role in describing observed behaviour of isolated skyrmions (diameter < 100nm). In the latter the region in which the magnetization rotates is measured to be around 30 nm. Such quantitative information on the multiscale magnetisation variations in the system is key to understanding and exploiting the behaviour of skyrmions for future device applications.

Magnetic skyrmions are 2D topological spin textures of significant current interest having been predicted to occur in materials with non-centro-symmetric crystal structures, resulting in a strong Dzyaloshinskii-Moriya interaction (DMI) [1,2,3,4,5]. The existence of such skyrmions is dependent on both field and temperature and mostly has been observed below room temperature. However interest has also focused on materials with a large spin-orbit coupling between heavy metal (HM) and ferromagnetic (FM) films which provides an interfacial DMI, in such a system skyrmions have been observed at low temperature [6]. In the case of thin FM layers the interfacial DMI results in chiral spin structures which have been proposed for both logic and data storage applications [7,8]. More recently, for ultra-thin magnetic layers with perpendicular magnetic anisotropy the presence of skyrmionic bubbles and skyrmion textures have been shown at room temperature [9,10,11,12]. Furthermore modification of the local film properties by means, for example, of focused ion irradiation have led to observation of controlled artificial configurations such as skyrmions and antiskyrmions [13]. HM/FM/HM multilayer stacks allow an engineering of the strength and sign of the interfacial DMI at both FM interfaces, and therefore control of the chirality of the spin texture. In particular, using HM/FM and FM/HM interfaces resulting in additive DMI which can lead to a large effective DMI and stabilize skyrmions at room temperature [11,12,14]. The dynamics of such skyrmions are of particular interest for possible applications using spin polarised current from both experimental and theoretical perspectives [12,14,15,16].

A key consideration in the study of these skyrmion systems is characterisation of the physical structure of the material system and, in particular, imaging of the magnetisation texture. Many microscopy methods have been used to investigate skyrmions including surface and transmission methods spanning a range of resolution to provide details on length scales from tens of nanometres down to atomic dimensions. Magnetic force microscopy (MFM) has been used to observe skyrmion states after magnetic field preparation or after pulse applications [15,17,18]. Spin-polarised low-energy electron microscopy (SPLEEM) allowed direct imaging of the chiral nature of domain or skyrmion walls together with a measure of the domain wall width (100 nm is quoted) for tuned systems with in-plane and out of plane magnetisation [19,20]. Imaging of domain/skyrmion size has also been carried out with magnetic transmission soft X-ray microscopy (MTXM) and scanning transmission X-ray microscopy (STXM). Here in-situ experiments have provided information on skyrmion size (~100 nm), formation and displacement using in-situ current pulses [12,15]. Using the technique of spin polarised scanning tunnelling electron microscopy (SP-STM) it has been possible to image at low temperature, compact skyrmions as small as a few lattice parameters, in Fe monolayers on Ir(111). Additionally control of their creation and annihilation has been demonstrated by passing current through the STM tip [6]. Lorentz microscopy is the study of magnetic structure (magnetic induction **B**) in the transmission electron microscope (TEM) and has been used to confirm the structure of Bloch skyrmions in B20 and Fe/Pd ML materials [2,21,22] and study chiral Néel walls in films with interfacial DMI [23,24].

In this paper we present a transmission electron microscope (TEM) imaging and analytical study of the physical, chemical and magnetic structure of a multilayer (Ir|Co|Pt) film system possessing perpendicular anisotropy and strong interface-driven DM interaction. The structural and chemical characterisations demonstrate the integrity of the multi-layer system and show that the physical structure of the film is consistent with recent dynamical creep-like behaviour of skyrmions [15]. Lorentz TEM (LTEM) is used to reveal details of the Néel character of the domain walls/skyrmions for repeat layer structures with additive DMI. Additionally, quantitative imaging has been carried out to characterise both the length scales of the domain periodicity, skyrmion size and domain wall width of these characteristic Néel-like structures. Measurement of the integrated magnetic induction confirms quantitatively that the skyrmionic magnetic structure is present in all layers, consistent with the presence of an inter-layer coupling in the PMA material system [25]. This study highlights the utility of all methods of TEM in studying these materials and the key quantitative information provided by imaging and analytical methods. We are able to show that LTEM images can be used to identify the wall texture and measure its spatial variation. LTEM is one of the few techniques which has the resolution (nanometre) to measure all length scales of magnetic texture in skyrmion and chiral materials and is also well suited to materials fabricated by methods suitable for device application.

Samples were prepared by room temperature dc magnetron sputtering deposition of the layered structure Pt(10)/[Ir(1)/Co(0.6)/Pt(1)]×*N*/Pt(3) where the numbers are thickness in nm and *N* the number of repeat layers studied, being 10 and 20 here. The interfaces of Ir/Co and Co/Pt have a DMI of opposite sign, however when arranged in this configuration either side of the thin Co layer the DMI from the two interfaces is additive [11,12,26,27]. The repetitions are used to increase the thermal stability of the structures and to increase the magnetic contrast observed in TEM. Similar samples have already been analysed using STXM showing sub-100 nm skyrmions stabilised by the large additive DMI present at the interfaces of the Pt|Co|Ir multilayers. This should guarantee the presence of Néel (hedgehog) skyrmions [11,12]. However, STXM is not sensitive to in plane magnetization and thus no information can be directly obtained on the domain wall type.

For TEM investigations, the film stacks were deposited on silicon nitride membrane windows which provide a 100×100 μm² square electron transparent substrate (~ 35 nm thick) suspended inside a 2 mm square silicon substrate (500 μm thick). This allowed for plan-view TEM experiments. Cross-sectional samples could be made from these films deposited on the thick silicon plus silicon nitride part of the substrates using focused ion beam preparation methods. Additional plan-view samples were also prepared on continuous thin carbon films on regular TEM grids. These substrates allowed for higher tilting of the samples in the TEM resulting in higher contrast from the PMA films. Magnetic characterisation of the samples was carried out using SQUID and alternating gradient field magnetometry. For reference, this allowed the 10 repeat ML sample properties to be estimated to be 0.96 ± 0.10 MA m$^{-1}$ for the saturation magnetization and an effective anisotropy of 0.17 ± 0.04 MJ m$^{-3}$. Such parameters served useful for micromagnetic simulations and to assist in quantification of the TEM results as will be discussed later in the paper.

Characterisation of the multilayer film system was carried out on a JEOL ARM200cF instrument operated at 200 kV. This allowed high resolution imaging and spectroscopic analysis of the multilayer structure from a cross-section sample with subnanometre resolution [28]. Electron energy loss spectroscopy (EELS) analysis meant that elemental composition could be determined and this utilised the Gatan Quantum 965ER spectrometer. The high resolution capability is possible due to a CEOS (probe) aberration corrector which is active in scanning TEM (STEM). As already stated, plan-view samples were prepared on silicon nitride membranes allowing both high angle annular dark field (HAADF) imaging to measure grain sizes and LTEM for imaging magnetic textures stabilized by the chiral DMI. In the case of the latter, the instrument was run in field free mode with the objective lens off or weakly excited to provide an applied field at the sample [29,30]. As the multilayers in this study have PMA, they need to be tilted to give a component of magnetic induction perpendicular to the electron beam. We initially used the defocused Fresnel imaging method to generate magnetic contrast as the beam is deflected in different directions from the magnetised regions. In consequence, the contrast can be interpreted to identify the domain wall type as either Néel or Bloch walls [23]. However in this study we also used differential phase contrast (DPC) in STEM with a pixelated detector for quantitative imaging of the induction distribution within the material [30]. DPC is an in-focus method and can image magnetic textures with a resolution down to 1 nm [29]. The pixelated detector allows a much more efficient means of measuring the deflection of the beam compared to standard DPC imaging with a quadrant detector, with these samples having a small magnetic deflection (of the order of microradians), such imaging is essential for quantitative imaging.

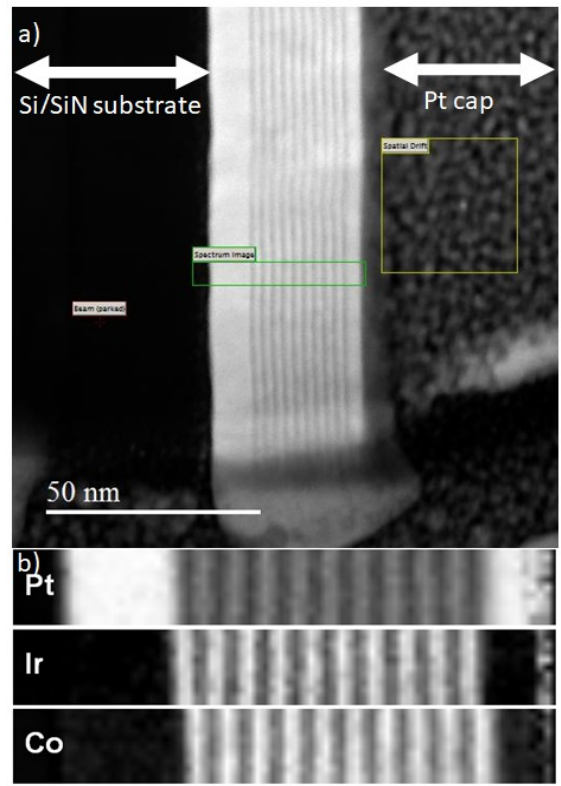

FIG. 1 Analytical electron microscopy study of a cross-section of a multilayer film. (a) High angle annular dark field (HAADF) image of cross-section showing multilayer structure (Pt/[Ir/Co/Pt]×10/Pt) between silicon substrate and a thick protective Pt cap layer deposited in the FIB before sectioning. The region where the EELS data was acquired is indicated by the green rectangle. (The yellow rectangular area was used during EELS acquisition to correct for spatial drift). (b) Elemental distribution of Pt, Ir and Co in the region (green rectangle in (a)) showing well defined layered structure.

The cross-sectioned sample was firstly imaged in high resolution HAADF mode. This is shown in Fig 1(a) for a 10× repeat multilayer sample. HAADF imaging is sensitive to atomic number ($Z$) of the material so images with low $Z$ appear darker than those with high $Z$. There are three regions clearly visible in the image: the dark region on the left is the silicon/silicon nitride substrate, the FM/HM multilayer (ML) in the centre and finally the thick protective ion beam deposited platinum cap on the right. In the multilayer region the Pt/Ir ($Z$=77, 78) appear equally bright and the Co ($Z$=27) layers appear darker. The 10 individual Co layers are shown as well defined and continuous in the region imaged. EELS analysis was carried out on the multilayer region to further demonstrate the integrity of the layer structure and composition maps for the Pt, Ir, and Co are shown in Fig 1(b). It should be noted that the pixel spacing for the elemental maps was 0.5 nm, which is close to dimension of the thinnest/narrowest layers here. The thickness of the cross-section itself was of the order of 60-80 nm as determined from the EELS measurements.

The plan-view sample was also imaged with high resolution to determine the grain size distribution in the film stack. Images were taken using dark field imaging in STEM. A typical image is shown in Fig 2(a), which shows the polycrystalline structure of the stack with a variable grain size that looks to be mostly < 10 nm in size. In dark field imaging the grains that appear dark are not well oriented for Bragg scattering and analysis is performed by thresholding the contrast and measuring the size of the dark grains

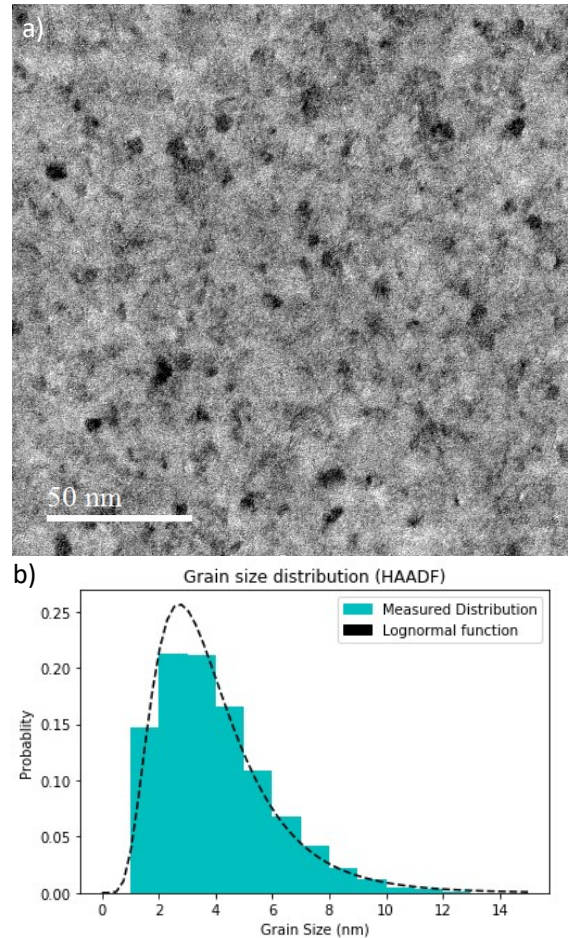

FIG 2. TEM dark field image and grain size distribution. (a) Dark field image showing grain structure in 10× multilayer repeat stack. (b) Histogram of grain sizes with log-normal fitting.

from a series of 30 images to provide reasonable statistics [31]. Overall this means around 10,000 grains are measured. The result of such an analysis are shown as a histogram of grain size in Fig. 2(b) together with a best fit log-normal distribution as standard [32]. The fitted log-normal function gives the mean grain size as 4 nm with a standard deviation value of 0.5 nm. The grain size may be a key factor in explaining some skyrmion motion. It was recently demonstrated that the granular structure may result in local variations of the magnetic properties and as such controls a creep-like regime at low current density/low velocity of the current-induced skyrmion motion [15]. The grain size is

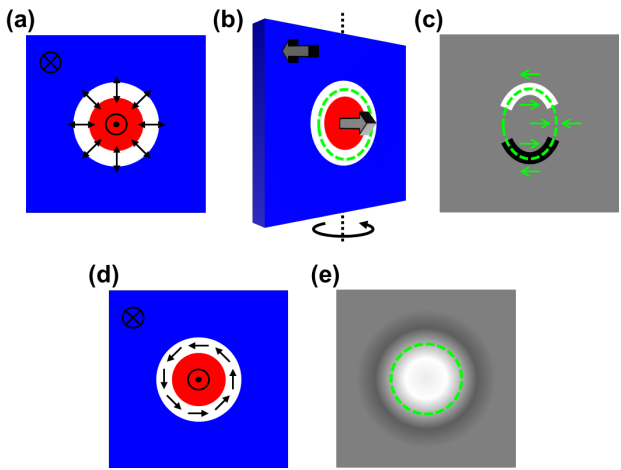

FIG. 3 (colour online). (a) Schematic representation of bubble domain/skyrmion in a thin film with perpendicular magnetisation. (b) Effect of tilting sample showing a finite magnetisation component perpendicular to the electron beam in the TEM. (c) Schematic of expected Fresnel contrast visible resulting from sample tilt. The green arrows show the magnetisation component orthogonal to the electron beam. (d) Magnetic configuration for a Bloch skyrmion. (e) Expected Fresnel contrast for the electron beam at normal incidence to a Bloch skyrmion.

expected to exert a greater influence on skyrmion motion when it approaches the size of the skyrmions. It is expected that the dynamics of the skyrmions will be less affected for such a grain size distribution here as the skyrmion size is more than an order of magnitude greater than the mean grain size. However it is worth noting that some larger grains do exist in any distribution and could play a role in pinning, hence control of the grain size is therefore of importance significant consideration in the materials analysis and design.

Chiral Néel DWs and/or skyrmion walls were first imaged in the defocused Fresnel mode of LTEM. Normally for in-plane magnetised thin film samples, domain walls appear as black and white lines due the deflection of the electron beam arising from the Lorentz force causing the beam to diverge or converge either side of the wall. However, in the case of domain walls in films with PMA, at normal incidence the magnetisation in the domains causes no deflection of the beam and in this orientation only the magnetization component within the domain wall itself may result in contrast. It has been shown that in such cases only Bloch walls may give contrast whilst Néel walls are invisible [23]. For the current samples, with the strong interfacial DMI present, the walls are expected to be of Néel type and we explain what to expect from the TEM imaging. For the multilayers this assumes the layers are all magnetised with the same configuration due to the strong inter-layer coupling. Figure 3(a) shows the configuration of the magnetization (all layers projected on to one) associated with a circular domain wall/skyrmion having in-plane magnetisation separating two domains with out of plane magnetisation. As stated, no contrast is generated assuming the electron beam is travelling in a direction perpendicular to the page. This is in part due the out of plane component of induction from the domains which thus gives no deflection, being directed parallel to the beam but also because of the divergent magnetisation component in the wall where the demagnetising field cancels the magnetisation projected along the beam [23,24].

Contrast can be generated by introducing a component of magnetic induction from the domains which is perpendicular to the electron beam by tilting the sample with respect to the beam as shown in Figure 3(b). This creates a magnetization configuration presented to the beam which causes it to diverge/converge at certain areas of the domain wall resulting in contrast shown schematically in Figure 3(c). It should be noted that contrast (in black and white) is only observed when the magnetization has a component with some aspect parallel to the length of wall. Where the magnetization appears head-to-head or tail-to-tail, no contrast is observed. The reason for this is again the fact that the magnetisation is divergent here, like the wall component at normal incidence. As such the wall appears discontinuous and the nature of the wall contrast also indicates the tilt axis, no contrast is observed where the wall runs parallel to the tilt axis. Note however that argument for contrast here is based only on the out-of-plane component of magnetisation, and such Lorentz imaging does not reveal directly the chirality sign of the domain wall. In the case of a Bloch skyrmion shown in Fig 3(d), the collective effect of the DW on the electron beam produces a light or dark (depending on the focus and chirality) spot even in normal incidence, as has been observed in B20 materials [2,5,21]. Therefore, it is clear that LTEM provides a clear distinction between Bloch and Néel skyrmion structures for film possessing PMA.

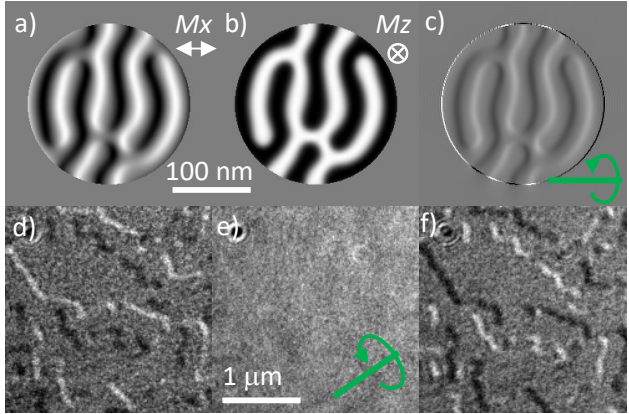

FIG. 4 A micromagnetic simulation of a single 0.6 nm Co layer film with interfacial DMI showing the (a) $M_x$ and (b) $M_z$ components of magnetisation. (c) Resulting simulated Fresnel image with beam tilt of 45°, tilt axis indicated in green. (d-f) show experimental Fresnel images of N = 10× ML sample taken at different sample tilt +30°, 0° and -30° respectively. The tilt axis is as indicated by the arrow.

A further illustration is made by a micromagnetic simulation performed using the MuMax3 program [33]. The simulation was carried out using material parameters $M_s = 1.1\times10^6$ A m$^{-1}$, $A_{ex} = 1.6\times10^{-11}$ J m$^{-1}$, $K_u = 1\times10^6$ J m$^{-3}$, $D = -5$ mJ m$^{-2}$ for a cylinder of diameter 200 nm and thickness 0.6 nm. The large $D$ value is not intended to mimic any experiment but was used to provide a multi-maze like domain structure with many walls to illustrate what might be expected experimentally. The voxel size used was 1 nm×1 nm×0.6 nm. The relaxed state is shown in Figure 4(a) and (b) with in-plane magnetisation component perpendicular to the predominant wall length ($M_x$) and the out of plane component ($M_z$). (The $M_y$ component is not shown.) The simulation confirms that the walls present are chiral and of the Néel type. A calculation of the expected Fresnel image for a beam tilt of 45° is shown in Figure 4(c). The contrast is as expected from the schematic in Figure 3(c) with black and white lines at the wall positions disappearing where the projected magnetisation appears head to head or tail to tail. It should be noted again that the contrast in the image is dominated by that from the magnetization component normal to the film, $M_z$.

Experimental Lorentz imaging of the films in the microscope was carried out at an accelerating voltage of 200 kV. Note that at normal incidence the total metal layer thickness is 39/65 nm for the 10×/20× repeat layers with a total magnetic layer thickness of 6/12 nm. At normal incidence, the magnetisation causes no deflection of the beam, only by tilting away from the normal by an angle, $\theta$, giving a component of magnetization $M_s\times\sin\theta$ ($M_s$ the saturation magnetisation) perpendicular to the beam, a signal appears. This tilt also means that the beam traverses a thicker total amount of material $t/\cos\theta$, for a film of thickness $t$. The large overall thickness and small component of magnetization means that the magnetic contrast is rather weak and a large defocus is required (1-10 mm). To prove that the walls from the experimental films are indeed of the Néel type we show a series of images taken at different tilts for a 10× ML film grown on a carbon film. These are shown in Figure 4(d-f) and represent the same area of film with tilt of +30°, 0° and -30°, the defocus used was 2 mm. In the same way that the Neel wall type was confirmed for a single layer [23] we can see that the contrast here is reversed at the ± tilt angles and there is no magnetic contrast in the untilted image. If the wall was of the Bloch type then strong contrast would be observed from the wall component at zero tilt as shown in Figure 3(e). We are also able to provide an estimate for the domain wall width from the Fresnel images. The width of the divergent (black) wall from the Fresnel image is normally much wider than the actual width due to the defocus used. By taking a series of images at different defocus and measuring the measured width variation, the wall width can be estimated from the value extrapolated to zero defocus for divergent wall contrast [34] (see supplementary information S1). With this method the domain wall width was measured to be 27 ± 8 nm (see supplementary information for further details). We return to this value when measuring by the in-focus DPC method later in the paper. Direct measurements of Néel domain wall width has been made using SPLEEM, although that was on a single crystal system with thicker Fe/Ni having in-plane magnetisation where the width was reported as 110 nm [20].

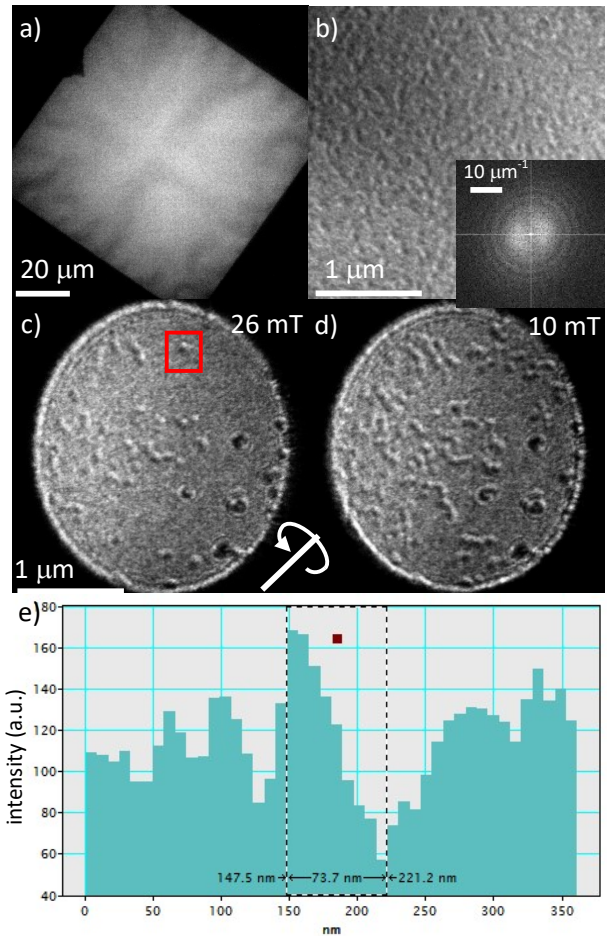

FIG. 5. (a) Low magnification in-focus image of Pt(10)/[Ir(1)/Co(0.6)//Pt(1)]×20/Pt(3) film, thickness in nm. (b) Tilted (20°) Fresnel image in demagnetised state showing high density of domain with inset of FFT from the image. (c) and (d) are tilted (20°) Fresnel images showing formation of skyrmion structure on reduction from saturating field which is applied along the beam direction. The tilt axis is indicated by the line and rotation direction. (e) Is a line trace (along the rotation axis) from the individual skyrmion in (c) indicated by the red box.

Samples grown on the silicon nitride membranes showed signs of stress as can be seen from an in-focus low magnification TEM image of the whole 100 μm×100 μm in Figure 5(a) for a 20× ML. The uneven contrast from the films results from buckling of the film noticeable at the edges but extending to the centre of the membrane. At higher magnification a defocused Fresnel image is shown from a tilted (20°) film in Figure 5(b). The contrast is rather low and the magnetic contrast is on a much smaller scale in this nearly demagnetised state. This is consistent with previous studies of films in similar systems by MTXM and MFM have shown that the domain sizes are of the order 100-120 nm [11,12]. In order to quantify the periodicity of these structures we analysed images by taking a Fourier transform which is shown in the inset of Figure 5(b). The transform here shows a series of rings which are associated with a loss of information at different spatial frequencies due to defocusing, effectively due to the transfer function of the microscope in this imaging mode. However also visible in the transform are diffuse lobes which are associated with the domain wall contrast. These lobes give a measure of the spread of domain size/periodicity present and for the 10x and 20x layers the average domain sizes are 120±20 nm and 125±20 nm respectively. These results are in excellent agreement with the STXM measurements [12] and are consistent with an expected value for $D$ of around 2.0 ± 0.3 mJ m$^{-2}$, calculated from this domain size [11]. It should be noted that the TEM images represent a projected image though all the layers and so this suggests that each layer has the same magnetisation configuration as was seen in the STXM images. Additionally field application from a uniform state shows that small scale skyrmion domains nucleate in the film as shown in Figures 5 (c) and (d). Here small individual skyrmions are visible on a scale of ~75 nm diameter. Again this is similar, though a little smaller, to that also seen by LTEM but in a different Co/Pd system where a value of ~ 90 nm is quoted [24].

Further quantification of the magnetic state of the films was provided by DPC using a pixelated detector [30]. This technique is practised in STEM and images obtained correspond to maps of integrated induction components. We use the STEM in field free mode and utilise a fast pixelated detector known as Medipix3. This allows us to record an image of the unscattered electron disk at each point in the scan, processing of the data then gives a sensitive induction profile of the magnetic structure. It should again be stated that these films are quite challenging for this technique due to the large thickness of film traversed by the beam and the low magnetic contrast resulting from the PMA material. A DPC image obtained at low magnification from a 10× ML films tilted at 45° using the pixelated detector is shown in Figure 6(a). We note that using a conventional quadrant DPC detector no discernible magnetic contrast was visible in this sample. The main reason for this is the large and non-magnetic signal variation arising from differential scattering from the crystallites. This is exacerbated due to the tilting of sample, resulting in the beam traversing a large overall film thickness relative to the magnetic layer thickness. The measured signal here from DPC is the deflection angle of the beam, which is linearly related to the integrated magnetic induction. In this case the deflection signal measured for the induction component perpendicular to the tilt axis is ±4 μrad. In the case of a film tilted at an angle $\theta$ the integrated induction for this component, as explained earlier, would be $B_s t \times \tan\theta$ which is just $B_s t$ for a tilt angle of 45°. Note that for a uniformly magnetised film with saturation magnetisation $M_s$ the saturation induction is just $B_s = \mu_0 M_s$. The deflection angle can be converted to an integrated field to give a value of ±6.6 T nm (±$B_s t$). For the 10× ML film we have a magnetic film thickness 10×0.6 = 6 nm of Co and this would then be consistent with a value $B_s$ = 1.1 T. From the magnetometry measurements given earlier for the Co the saturation magnetisation was quoted as 0.96 ± 0.10 MA/m which is equivalent to Bs = 1.12 T. Therefore, it has been proven quantitatively that all the layers have the same magnetic structure as expected from the inter-layer coupling. Previous studies using MTXM have deduced such homogeneity in the layers qualitatively through domain regions showing as only black and white and lacking levels of gray [11,12].

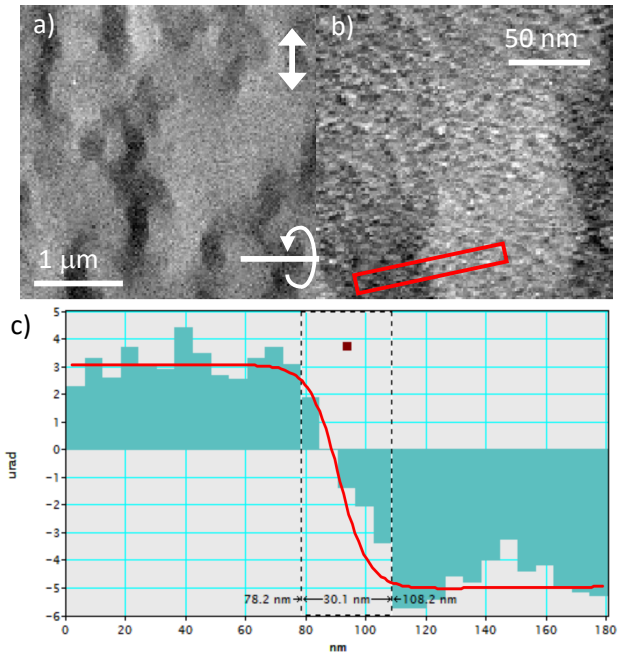

FIG. 6. DPC images taken at (a) low and (b) high magnification from the 10× ML film tilted at 45° to the electron beam. The tilt axis is shown in (a) together with the mapping direction of the magnetic induction. A linetrace taken from the red box area in (b) is shown in (c) averaged over 5 lines showing the deflection (integrated induction) profile allowing measurement of domain wall width. Also shown in red is the profile obtained from the MuMax3 simulation, see supplementary information.

We further imaged the domain wall structure at higher magnification as shown in the DPC image in Figure 6(b). The same component of magnetic induction is mapped as in Figure 6(a) and is parallel to the domain wall. It can be seen that even with the pixelated detector method some residual contrast associated with crystallite structure is still present in the images. Again this is due to the scattering from the crystallites in addition here we have the large overall film thickness, which results in a noisy signal at the detector. However at this magnification we are able to take a linetrace across a domain wall and this is shown in Figure 6(c). For this wall we see that the wall width is measured as 30±5 nm. MuMax3 simulations for an isolated wall in this material indicate that such a wall width is consistent with parameters given for the film earlier (see supplementary information S2). This value is consistent with that measured indirectly using the focus series with Fresnel imaging and extrapolation given earlier in this paper. Most of the imaging methods described earlier do not have resolution to directly image the wall width apart from SPSTM which has been performed on monolayer single crystal structures [10]. One example of imaging walls in Co/Pt MLs is that of X-ray holography where image processing is used to obtain a domain wall width also of 30±10 nm, although the resolution for the technique is noted to be 30nm [35].

Conclusions

This TEM study reveals crucial structural parameters: continuity of the magnetic layers throughout the full structures and the distribution of grains sizes peaking around 4 nm in diameter. This information is important to understand pinning and creep mechanisms in sputtered structures. Lorentz imaging confirmed the Néel character of the domain walls in these Ir|Co|Pt structures where large interfacial Dzyaloshinskii-Moriya interaction magnitude were measured. LTEM allows lengthscale measurements of the key magnetic quantities present in these films namely, domain size, wall width and skyrmion extent ranging from > 100 nm down to a few tens of nm. Additionally the domain wall width has been quantified using DPC imaging.

It has been shown here that the Néel character of the chiral magnetisation has been demonstrated by the lack of LTEM contrast from the divergent Néel wall component. However there does exist a possibility to reveal directly information about this component and hence deduce the sign of the DMI from LTEM. This is explored in the supplementary information (S3) whereby image calculations of wall structures are made for simulated one dimensional wall with the beam at oblique angles of incidence. These calculations demonstrate a clear signature that shows this component could potentially be imaged. So far we have been unable to see such a signature in experimental images where background signal variations are present and the presence of scattering from crystallites also limits the sensitivity of what may be observed. An important point is that the calculations are based on the assumption of an identical structure in every magnetic layer of the stack. However it has recently been reported that surface states may be present in ML systems where the Néel components are of opposite signs at the top and bottom layers of the stack [36]. If such a configuration were present in the MLs here the divergent Néel wall component would likely remain invisible in LTEM, even at oblique incidence.

The functionality of skyrmions is being realised by the intense effort worldwide in this area, studies providing high resolution information on the physical, chemical and magnetic structure are vital to our understanding of these complex magnetic phenomena.

## Methods

### Sample preparation

The thin films were deposited by dc magnetron sputtering in a vacuum system with a base pressure of $2\times10^{-8}$ mbar. An argon gas pressure of 6.7 mbar was used during the deposition. The target and the substrate were separated by roughly 7 cm. Typical Co growth rates of around 0.1 nm s$^{-1}$ were achieved.

### TEM imaging

All the TEM imaging shown here was carried out on a JEOL ARM 200cF equipped with a cold field emission gun and CEOS probe aberration corrector. EELS and HAADF images were acquired with a 40 μm condenser 1 aperture, spot size 5 and a 2 cm camera length resulting in a convergence angle of 36 mrad and collection angle at the detector of 29 mrad. For these conditions the probe current was 180 pA and probe size < 0.2 nm, which is well below the sampling pixel size of 0.5 nm used in this paper. LTEM was performed in both TEM and STEM modes. In the case of Fresnel imaging a lens defocus of between 1 and 10 mm was used. In STEM DPC imaging used a condenser 1 aperture of 10 μm which gives a probe of 7.0 nm and a resolution of 3.5 nm, with a sampling pixel size of 5.9 nm at the highest magnification used here.

### Micromagnetic simulations

MuMax3 simulations for a domain wall in single magnetic layer thickness 0.4 nm and pixel size of 2 nm in-plane for an area of 400 nm by 300 nm. A single domain wall was simulated as a starting state with material parameters $A_{ex}$ = 16 pJ m$^{-1}$, $D$ = 2.0 mJ m$^{-2}$, $K_u$ = 0.9 MJ m$^{-3}$ and $M_s$ = 1.1 MA m$^{-1}$.

**Acknowledgements** :
The authors acknowledge financial support from European Union grant MAGicSky No. FET-Open-665095 and EPSRC through grant EP/M024423/1.

Author contributions
N.R., V.C. and A.F. conceived the project and the material design. W.L., S.C., K.G. and K.Z. grew the films. S. McV. planned the TEM/STEM experiments and together with S. H., K. F., S. McF., D. McG. and M. K. acquired and analysed the experiments and data. S. H. and K. F. performed the micromagnetic simulations and TEM image calculations. S. McV., N. R., V.C., K. Z. and C. H. M. prepared the manuscript. All authors discussed and commented on the manuscript.